\documentstyle[aps,prd,preprint,tighten,floats,epsf]{revtex}

\begin{document}
%

\preprint{VAND-TH-98-14}

\title{GUT COSMIC MAGNETIC FIELDS IN A WARM INFLATIONARY UNIVERSE}

\author{Arjun Berera, Thomas W. Kephart, and Stuart D. Wick}

\address{
   Department of Physics and Astronomy,
   Vanderbilt University,
   Nashville, TN 37235, U.S.A.
   August 1998
}


\maketitle

\begin{abstract}
Sources of magnetic fields from grand unified theories are studied in the
warm inflation regime.  A ferromagnetic Savvidy vacuum scenario is
presented that yields observationally interesting large scale magnetic
fields. As an intermediate step, a general analysis is made of defect
production at the onset of warm inflation and monopole constraints are
obtained.  Many features of this Savvidy vacuum scenario are applicable
within a supercooled inflation regime and these points are discuused.

\vspace{0.34cm}
\end{abstract}

\medskip

hep-ph/9809404

\medskip

In Press Physical Review D 1998

\medskip
\medskip


\medskip

\bigskip


\bigskip


\section{Introduction}
\label{sec.int}

Although magnetic fields make up only a very small fraction of the
energy density of the Universe, they make a profound impact on what we
observe. They influence star and galaxy formation and evolution, they
are responsible for diverse effects, from altering atomic spectra
to the general lack of hydrostatic equilibrium in a plasma.

There is sufficient ionization in the galaxy that the galactic B-fields
are locked into and flow as a fluid with the plasma. Given any
contour $C$ embedded in the plasma at time $t$, then at a later time
$t'$ the contour will have moved along with the plasma to $C'$. To a
very good approximation the flux through $C$ is equal to the flux
through $C'$. The relaxation time for the B-fields in 
our galaxy $\tau$ is approximately \cite{kronberg,parker} $L^2 \sigma/c^2$ where 
$L$ is the
characteristic length scale of the magnetic fields (or coherence length
of the plasma) and $\sigma $ is the conductivity. For $L$ a few hundred
pc and typically one free electron per $cm^3$ contributing to the
conductivity, $\tau$ is many orders of magnitude longer than the age of
the Universe. Likewise the relaxation time for extra-galactic magnetic
fields is also expected to be much longer than the age of the Universe.
Typical galactic B-fields are $\sim 10^{-6} {\rm G}$ while extragalactic
magnetic fields are thought to be generally less than 
$\sim 10^{-11} {\rm G}$,
although in galactic sheets there is evidence for fields as strong as
$\sim 10^{-7} {\rm G}$ \cite{biermann1}.

The mechanism for the generation of cosmic magnetic seed fields is not
known, although several possibilities have been suggested:

(1) Gauge theories may have ferromagnetic vacua,
\cite{savvidy},
often referred to as Savvidy vacua, which
lead to nonvanishing B-fields imprinted on the pregalaxy formation
plasma \cite{eo1}.
This method can work if the $\beta$-function for a gauge theory
satisfies a plausible integration condition so that the effective
Lagrangian has a minimum $g^2 [TrF^2]_{min} \sim \Lambda^4$, away from the
perturbative ground state $TrF^2 = 0$, where $\Lambda$ is the RG scale.
In particular, the requirement on the $\beta$-function is
\begin{equation}
\left| \int_g^{\infty} \frac{dt}{\beta(t)} \right| < \infty.
\label{betaf}
\end{equation}
One way to realize the nonperturbative minimum is with a constant
B-field $B \sim F$ so that $g^2B^2 \sim \Lambda^4$. The one-loop (zero
temperature) effective potential in an SU(N) background 
B-field is \cite{savvidy,masa}:
\begin{equation}
V(B) = B^2/2 + \frac{11N}{96\pi^2}g^2B^2(\ln\frac{gB}{\mu^2}
-\frac{1}{2})
\label{effpotb}
\end{equation}
minimizing this expression gives
\begin{equation}
g B_{min} \equiv \Lambda^2_{\rm SU(N)}
= \mu^2 \exp{\left( \frac{-48\pi^2}{11Ng^2} \right)}
\label{bmin}
\end{equation}
and
\begin{equation}
V(B_{min}) \sim -(gB_{min})^2 
\left( \frac{11N}{192 \pi^2} \right) 
\end{equation}
In the early Universe there are thermal corrections to $V(B)$, but it
can be argued \cite{eo1} that these corrections are small enough to allow the
$B_{min} \neq 0$ vacuum to exist at all $T$. 
In the Enqvist-Olesen scenario \cite{eo1}, the transition to the
$B_{min} \neq 0$ phase is caused by local fluctuations due to
currents in the plasma in the very early Universe. At the GUT scale, 
they show the possibility
to have horizon-size persistent quark currents that lead to
magnetic fields $B \sim T^2$, which in turn precipitate the phase
transition to the $B_{min} \neq 0$ vacuum. This leads to a B-field of 
approximately ${10}^{42} {\rm G}$ at the GUT scale
which would evolve to a $\sim {10}^{-14} {\rm G}$ field today in the absence of
dynamo effects.

(2) Gradients in the VEVs of Higgs fields at cosmological phase
transitions can lead to B-fields on completion of the phase transition
\cite{tv}. To summarize this method we consider a gauge theory with
gauge group $G.$  If a Higgs field $\phi$ develops a
vacuum expectation value (VEV) $\langle \phi \rangle \neq 0$, then the theory
undergoes a phase transition and $G$ breaks to some subgroup $H$. 
$\langle \phi \rangle$ 
will be correlated up to some length scale $\xi_i$ that can be the
horizon scale for a second order phase transition, but is in general
smaller for a first order phase transition.  The fact that 
$ \langle \phi \rangle$ has
a maximum correlation length implies $ \langle \phi \rangle$ is 
non-uniform over length scales larger than $\xi_i$ and therefore
$\partial_{\mu}  \langle \phi \rangle \neq 0$. 
Furthermore, it can be argued that the
variation of $ \langle \phi \rangle$ can not be compensated with a gauge
transformation over length scales larger than the horizon size.
Therefore, 
$ D_{\mu} \langle \phi \rangle = 
(\partial_{\mu} -ieA_{\mu}) \langle \phi \rangle \neq 0$.
For the phase transition taking place at temperature $T_c$ one estimates
$\xi_c = (g T_c)^{-1}$ where $g$ is the gauge coupling constant; hence
$ \partial_{\mu} \langle \phi \rangle \sim  
g A_{\mu} \langle \phi \rangle \sim \langle \phi \rangle  \xi_c^{-1}$ 
and so $B \sim F \sim (g {\xi}_c^2)^{-1} \sim gT_c^2$.
For the electroweak phase transition $T_c \sim {10}^2 {\rm GeV}$ one finds a
magnetic field $B \sim {10}^{23} {\rm G}$ which then evolves with the expansion
of the Universe to provide predynamo seed 
fields $B \sim {10}^{-19} {\rm G}$
\cite{kronberg,eo1,tv}.

Both scenarios (1) and (2) give only small seed fields that must
subsequently be amplified by dynamo effects.

(3) The very early Universe may have been devoid of magnetic fields, but
seed B-fields could have formed in stars. These fields could then be
amplified by dynamo effects and expelled into interstellar space.
Interstellar B-fields could then be amplified again by galactic dynamo effects
to give the fields we see today \cite{biermann2}.

Scenarios (1) and (2) are both top down in the sense that they generate
seed B-fields that are trapped and amplified in galaxies and stars. As
these fields are generated very early, typically between the GUT scale
and the electroweak scale, they are cosmology dependent.  (3) is bottom
up as the fields are generated on small (stellar) scales and then
expelled into interstellar and subsequencely intergalactic space.
Consequently, (3) is rather insensitive to the cosmological model being
employed.

In this paper cosmic magnetic field generation will be examined during
warm inflation, based on the ferromagnetic vacuum mechanism (1) and the
Higgs gradient mechanism (2).  A scenario will be presented for
generation of cosmologically interesting magnetic fields during a GUT
symmetry breaking warm inflation.  
As a review, section II discusses
coherent magnetic fields during generic 
GUT and electroweak phase
transitions and section III summarizes warm inflation.
The main analysis of magnetic fields during warm inflation is in section
IV.  Subsection IVA examines the Higgs gradient mechanism. As an 
aside, from this analysis monopole constraints are 
derived for warm inflation.  Then in
subsection IVB the ferromagnetic Savvidy vacuum scenario is presented.  
Also discussed here are several features about this Savvidy vacuum
scenario that are applicable within a supercooled inflation regime.
Finally the results are summarized and scrutinized in the conclusion.

\section{GUT Magnetic Fields and Symmetry Breaking}

To have a specific model exemplifying the results which follow, this paper
will consider minimal SU(5). For this model,
$SU(5)$ gauge symmetry exists at high temperature above a critical temperature
$T_c.$  GUT symmetry breaks in phase transitions
via a pattern of spontaneous symmetry breaking (SSB)
to the low energy vacuum, $SU(3)_{C} \times U(1)_{EM}.$  In minimal
$SU(5)$ the breaking pattern is  
$SU(5) \stackrel{<\bf{24}>}{\longrightarrow} SU(3)_{C} \times SU(2)_{L} 
\times U(1)_{Y} \stackrel{<\bf{5}>}{\longrightarrow} SU(3)_{C} 
\times U(1)_{EM}$ 
with symmetry breaking scales $T_{c}^{GUT} \sim 10^{15}{\rm GeV}$, 
$T_{c}^{EW} \sim 10^{2} {\rm GeV}$, followed by quark confinement at
$T_{c}^{QCD} \sim 10^{-1} {\rm GeV}$. 
Cosmic magnetic fields generated at or near the GUT phase transition
will need to survive until the era of galaxy formation 
with sufficient strength to 
be candidate dynamo seed fields.  We review 
the $SU(5)$ Yang-Mills Lagrangian, identify the magnetic fields, and
subsequently consider the effect of SSB upon those fields.   
 
The Yang-Mills term in the $SU(5)$ symmetric Lagrangian density is
\begin{equation}
{\cal{L}}_{YM}= -\frac{1}{2} Tr(F_{\mu \nu} F^{\mu \nu})
\end{equation}
where
\begin{equation}
F_{\mu \nu} = \tau^{a}F^{a}_{\mu \nu},  
\end{equation}
\begin{equation}
F^{a}_{\mu \nu}= \partial_{\mu}A^{a}_{\nu} 
- \partial_{\nu}A^{a}_{\mu} + i g f^{abc}A^{b}_{\mu}A^{c}_{\nu},
\end{equation}
$a (=1,...,24)$ is the group index, 
$\tau^{a}$ are the 24 generators of
$SU(5),$ $f^{abc}$ are the structure constants, and
$ A^{a}_{\mu} $ are the gauge fields with coupling $g$.
The magnetic fields are given by
$B^{a}_{i}=\epsilon_{ijk}F^{a}_{jk}.$  
The fields $B^{a}$ are associated with massless gauge fields, but 
during SSB some gauge fields acquire a mass screening their 
respective components of the magnetic field.
The long range fields after the GUT phase transition correspond to 
the $ SU(3)_{C} \times SU(2)_{L} \times U(1)_{Y}$
generators that satisfy $\tau^{b} <\phi_{24}> = 0,$ where $\phi_{24}$ is
the Higgs field configuration responsible for GUT symmetry breaking.  

Magnetic fields could be established prior to SSB,
as in scenario (1) which we show applies to warm inflation.
What effect will SSB, originating in the Higgs sector,
have upon coherent B-fields of the Yang-Mills sector?  
To present a simple analysis of this effect 
suppose the $SU(5)$ symmetric theory is endowed with a magnetic field   
\begin{equation}
\sqrt{B^{a}B^{a}}= B_{GUT} \; (a=1,...,24),
\end{equation} 
where $B_{GUT}=B_{\rm min},$ in scenario (1).
Following the GUT phase transition half of the gauge
fields, the $X$ and $Y$ bosons, become massive and screen
magnetic fields.  The field strength then becomes
\begin{equation}
\sqrt{B^{b}B^{b}}
\sim \frac{1}{2}B_{GUT} \; \; \; (b=1,...,12)
\end{equation}
This result follows from the assumption that
$B_{GUT}$ is shared equally among
all massless field components.
By the same argument $B^{color}\sim \frac{1}{3} B_{GUT},
B^{weak} \sim \frac{1}{8} B_{GUT} ,$ and
$B^{Y} \sim \frac{1}{24} B_{GUT}.$  
                 
Similarly for the electroweak phase transition, we suppose
there to be a magnetic field immediately prior to SSB of strength  
\begin{equation}
\sqrt{B^{c}B^{c}} \sim B_{EW}, \; (c=1,...,4).
\end{equation}
These fields could originate from either scenario (1) or (2).
During SSB the $W$ and $Z$ bosons become massive and 
screen the weak magnetic field.  
The field strength is then 
$B^{em} \sim \frac{1}{4} B_{EW}.$
$B^{em}$ is obtained via the usual weak mixing angle
\begin{equation}
B^{em}_{i}= \epsilon_{ijk} F^{em}_{jk}
= \epsilon_{ijk} \left[ - \sin \theta_{W} \psi^{a}
F^{a}_{jk} + \cos \theta_{W} F^{Y}_{jk} \right]
\end{equation}
where 
\begin{equation}
\psi^{a} \equiv \frac{\psi^{\dagger}_{5} \sigma^{a}
\psi_{5}}{\psi^{\dagger}_{5}\psi_{5}},
\end{equation}
$\sigma^{a}$ are generators for the $SU(2)$ subgroup of $SU(5),$
and $\psi_{5}$ is the Higgs field configuration responsible for
electroweak SSB.  

$B^{em}$ could be the primordial seed field for galactic
dynamos and thereby ultimately responsible
for the large scale magnetic fields observed today.  
Based upon a counting of broken group generators, 
SSB alone reduces a GUT primordial magnetic field strength 
by a factor $\sim 24.$  Of course, evolution of the cosmic scale
factor until the time of galaxy formation will be primarily
responsible for magnetic field suppression.          

\section{Review of Warm Inflation}

Inflation in the presence of nonnegligible radiation is characterized by
non-isentropic expansion \cite{spbr,gmn,rudnei,ab2} and thermal seeds
of density perturbations \cite{mati,bf2}.  Such a region
can be realized as an
extension to supercooled inflation scenarios, in which reheating still
is necessary, or it can be realized in warm inflation scenarios \cite{wi}
in which there is no reheating.
In the warm inflation regime, the vacuum energy  density $\rho_v$
dominates the energy density, thus driving inflationary growth of the
scale factor.  In addition, there is a substantial component of radiation
energy density $\rho_r$. An analysis of two fluid Friedmann cosmology
\cite{ab2} composed of $\rho_v(t)$ and $\rho_r(t)$ shows that inflationary
expansion requires $\rho_r < \rho_v$ and, moreover, observationally
interesting expansion already occurs for $\rho_r \leq \rho_v/10$.  This is
a sizable radiation energy density component for most particle physics
early universe phenomena.  

In the context of scalar field dynamics, warm inflation can be realized by
a Ginzburg-Landau kinetic equation of motion
\begin{equation}
\eta {\dot \phi}= -\frac{dV}{d\phi}
\label{disseom} 
\end{equation}
Since ${\dot \rho_v(t)} ={\dot \phi} (dV/d{\phi})$, the dissipative term
$\eta {\dot \phi}^2$ expresses the loss of energy from the $\phi$-system
into the radiation system.
Such an equation of motion, for an SU(5) potential was shown to yield
an observationally consistent warm inflation scenario with respect to both
expansion e-folds, $N_e > 60$, and density perturbations
$\delta \rho/\rho \sim 10^{-5}$ \cite{wi}. 

The fundamental equation of motion for scalar fields is second order in
time. To obtain eq. (\ref{disseom}) from first principles, two steps are
necessary. First, a second order damped equation of motion must be derived
\begin{equation}
{\ddot \phi} + \eta(\phi) {\dot \phi} + V'({\phi}) = \xi(\phi,t),
\label{eq2order}
\end{equation}
where $\xi(\phi,t)$ is a random force term with
$\langle \xi(\phi,t) \rangle =0$.
Second, the validity of the first step must be demonstrated in the
overdamped limit $\eta (\phi) {\dot \phi} \gg {\ddot \phi}$
for the ensemble average of eq. (\ref{eq2order}).  With these two
steps, the resulting first principles equation of motion is of the form
eq. (\ref{disseom}).  The general consistency of this procedure with the
principles of quantum mechanics and warm inflation
cosmology has been demonstrated in a
quantum mechanical model in \cite{wi}.  
Based on methods developed in \cite{hs1,morikawa,GR},
eq. (\ref{disseom}) also has
been derived from first principles in a quantum field theory model in
\cite{bgr} for a symmetry restored $\lambda \phi^4$-model coupled to
scalar bosons. 

Here we will assume the general phenomenological validity of 
eq. (\ref{disseom}) and examine its consequences for generation of magnetic
fields during warm inflation.  A compact analysis of a variety of warm
inflation expansion behavior can be obtained from the class of potentials
\begin{equation}
V(\phi) = \lambda M^{4-n} (\phi - M)^n \; \;(\rm{symmetry}\;\rm{broken})
\label{wipotential1}
\end{equation} 
\begin{equation}
V(\phi) = \lambda M^{4-n} \phi^n \; \;(\rm{symmetry}\;\rm{restored}).
\label{wipotential2}
\end{equation} 
Exact warm inflation scale factor solutions for this class of potentials
are given in \cite{ab2}.  Here the results are summarized.  Warm inflation
scale factor trajectories, by definition, are required to start within a
radiation dominated regime $R(t) \sim \sqrt{t}$, go through
a positive acceleration behavior starting at $t_{BI}$ (begin inflation)
and then return to a radiation dominated regime at $t_{EI}$ (end
inflation).  During the course of this,  radiation energy is being
produced from vacuum decay and redshifted due to expansion, with a net
effect of gradual monotonic decrease in the density.  As such, in the warm
inflation regime there is no reheating.  In \cite{ab2} it is shown that
warm inflation scale factor solutions exist for potentials in
eqs. (\ref{wipotential1}) and (\ref{wipotential2}) when
$2 \leq n < 4$.  Within this range, for larger $n$, the duration of the
warm inflation period $t_{EI}-t_{BI}$ as well as the ratio of the initial
to final radiation energy density $\rho_r(t_{BI})/\rho_r(t_{EI})$ increase
for fixed e-folds $N_e=\ln[R(t_{EI})/R(t_{BI})]$.
The case $n=4$ yields an inflationary power-law expansion and sustains a
sizable radiation component.  However, for this potential the inflationary
period never terminates.

In a realistic setting, no single index $n$ describes the 
whole potential.
For example, given the scalar symmetry restored potential,
$m^2/2 \phi^2 + \lambda/24 \phi^4$, $n=4$ behavior at large $\phi$ goes
over to $n=2$ behavior as $\phi \rightarrow 0$.
Also in realistic models, the dissipative coefficient $\eta(\phi)$ may be
a function of $\phi$ \cite{bgr}.  For a simple power-law form
$\eta(\phi)=\eta_0 \phi^m$, the scale factor solutions for potentials with
index $n'$ will be the solutions for the potentials
in eq. (\ref{wipotential2}) for index $n=n'-m$ and dissipative
coefficient $\eta_0$ in eq. (\ref{disseom}).

For symmetry breaking warm inflation scenarios, it is important that the
inflaton's roll down the potential commences soon after vacuum
domination, so as to allow radiation production from vacuum decay.
This is necessary so that depletion of radiation energy density due to
expansion is compensated adequately by production. Thus
the inflaton should not get trapped in a metastable state once vacuum
domination occurs.  Second-order phase transitions are ideally suited
for this purpose.  Fluctuation induced first-order phase transitions
(``weakly first-order'') also can suffice if the decay rate from the
metastable minima at $\langle \phi \rangle =0$ is fast.  Aside from the
order of the transition the kinetics of the quench
during symmetry breaking is also important.  A complete treatment 
of order parameter
behavior near symmetry breaking is beyond the scope of this work.  It
should be noted that analyses of gauge-Higgs system transitions include
the range from first-order, weakly first-order and second-order
\cite{phaset}.

\section{Magnetic Fields During Warm Inflation}

Local magnetic fields can be produced by agitation of charge carriers. 
At high temperature thermal fluctuations can initiate such a disturbance when
$T \gg {\rm m}$, where ${\rm m}$ is a characteristic 
scale for the charge carrier. In
the particle physics phase transition scenario of the early universe, the
chronology of this condition with respect to the charge carriers is
equivalent to the chronology of the early universe.  As such, thermal
fluctuations are a generic source for production of magnetic fields in the
early universe.

Due to the nonnegligible thermal component before and during warm
inflation, magnetic field formation is apt to occur.  Production
mechanisms based on thermal fluctuations are random.  In general the
treatment of a random system requires identifying the random variables
and specifying the statistical distribution.  For warm inflation, two
mechanisms are suitable for a GUT-based scenario, the Higgs field
fluctuation mechanism of Vachaspati \cite{tv} and the ferromagnetic
Saviddy vacuum of Enqvist and Olesen \cite{eo1}.
These mechanisms originally were formulated for the post-inflationary
radiation dominated regime.  We observe that the warm inflation regime
provides suitable conditions for them to be active \footnote{
Some of these results were presented in \cite{vuconf}.}.

\subsection{Higgs Gradients - Monopole Constraint}

In the Vachaspati mechanism, the random sources for production of magnetic
fields are associated with the charged Higgs fields.  In the original
proposal, the Higgs field itself was taken as the random variable
with a coherence length $\xi \sim 1/m_H$. Here $m_H$ is the thermal
Higgs mass, which is assumed to be approximately equal to the vector 
boson mass, thus $\xi \sim 1/(gT)$ \cite{tv}.  A subsequent analysis
\cite{eo2} indicates that the appropriate random variables
are the gradients of the Higgs field.  The implication is, the
rms-magnetic field over $N_d$ independent coherent domains decreases as
\begin{equation}
B \sim \frac{gT^2}{\sqrt{N_d}}.
\label{brmsgen}
\end{equation}

The fluctuations of the Higgs gradients are not stable sources of magnetic
fields, however, near the critical temperature these fluctuations can get
locked or ``freeze-out''.  For the SU(5) transition these 
fluctuations will lock into topologically stable 
monopole and antimonopole configurations and these are the  
magnetic sources.  As such, a conflict of interest arises since
an overdensity of monopoles has inappropriate cosmological consequences, 
in particular the critical density constraint and the Parker bound limit.
These constraints in turn limit the size of magnetic fields
that are allowed from magnetic monopole sources.

To estimate these constraints, the number density of the Higgs
field fluctuations is required at the time of freeze-out.
The conventional estimate for the freeze-out temperature
is the Ginzburg temperature 
$T_F = T_G \stackrel{<}{\sim} T_c$.  $T_G$ is based on an equilibrium 
estimate
above which thermal fluctuations can still disorient correlated domains.
An alternative argument for freeze-out has been given by Zurek
\cite{zurek}, based on the kinetics of the quench as $T_c$ is approached
from above.  His analysis suggests that $T_F = T_Z > T_c$,
which is in qualitative
distinction to $T_G$.  Moreover, the number density of the defects,
$n_{\phi},$ depends on the kinetics.  In 1-D and 2-D Ginzburg-Landau 
models, similar in form to eq. (\ref{eq2order}) \cite{lagzur}, 
$n_{\phi}$ 
at $T_F=T_Z$ is shown to depend on the 
viscosity coefficient $\eta(\phi)$ and the
duration of the quench $\tau_Q$. In the overdamped regime
$n_{\phi} \sim (\eta/(m^2\tau_Q))^{1/4}$ and in the underdamped regime
$n_{\phi} \sim (1/(m \tau_Q))^{1/3}$,
where $m^2 = V''(0)$ is the characteristic dynamical
mass scale.  Although similar results have not
been confirmed for 3-D systems, the scaling is expected to hold
but the premultiplying coefficient should be smaller \cite{pablo}. 

For the warm inflation scenario, these findings lead to very different
possibilities.  Above $T_c$, when the scalar field is in the symmetry
restored regime, the viscosity coefficient has been shown 
\cite{bgr} in a 
$\lambda \phi^4$ quantum field theory model to vanish as
$\eta \sim \phi^2/T_c$.  This allows for the possibility of substantially
small defect production before entering the warm inflation regime.
For our present purposes, we will use the upper limits on defect density
production, since it is important to know the maximum monopole density
that warm inflation must dilute.   By either the Ginzburg or
Zurek criteria, the defect density should not exceed the intrinsic microscopic
scale, which at high temperature is set by $T \sim T_c$.   
Thus, we will consider the number density of monopoles at the onset
of warm inflation to be $n_M \sim T_c^3$.  Note the 
associated magnetic field energy density from eq. (\ref{brmsgen}),
$\rho_B \approx g^2T_c^4 \ll \rho_r \approx g^* T_c^4$,
where $g^*$ is the number of relativistic particles.

The critical density condition on monopoles requires \cite{kotu}
\begin{equation}
\Omega_M h^2 \approx 10^{24} (\frac{n_M}{s})(\frac{M}{10^{16}{\rm GeV}})
\leq 1
\label{critd}
\end{equation}
where $M \sim 4 \pi M_{GUT}/g^{2} \sim 20 M_{GUT}$ is the monopole mass. 
Ignoring monopole-antimonopole annihilation, the
ratio $n_M/s$ is constant \cite{kotu}, so it can be estimated at the end of
the warm
inflation, which yields
\begin{equation}
\frac{n_M}{s} = \frac{45}{2\pi^2 g^* e^{3N_e}} (\frac{T_c}{T_{EI}})^3
=\frac{1}{75 e^{3N_e}} (\frac{T_c}{T_{EI}})^3,
\label{nms}
\end{equation}
where the number of relativistic particles for
minimal SU(5) GUT is $g^* \sim 170$.
The temperature at the end of warm inflation $T_{EI}$ is required.
For the quadratic potential in eqs. (\ref{wipotential1})
and (\ref{wipotential2}) this is $T_{EI} \approx T_c/\sqrt{N_e}$
\cite{ab2}.
In general for the potentials in 
eqs. (\ref{wipotential1}) and (\ref{wipotential2}),
$T_{EI} \sim T_c N_e^{-p}$ for some small positive
exponent $p$.  The quadratic case is a good estimate, since near 
the minimum of a generic potential, when warm inflation ends, the behavior
is quadratic.  Based on eqs. (\ref{critd}) and (\ref{nms}) we find
$N_e \stackrel{>}{\sim} 20$.  The analysis of the Parker bound does not
change this lower bound.  This $N_e$ is sufficiently large that
an initial magnetic field as large as $\sim 10^{46}$G is miniscule
by the end of warm inflation.  A similar conclusion was arrived by Davis
and Dimopoulos \cite{dd}, who examined the Vachaspati mechanism during a
false vacuum inflation.  

The magnetic monopole constraints can be
removed by going to flipped-SU(5). In this model stable monopole
solutions are not admitted, although unstable
monopole-antimonopole configurations attached by strings 
are possible \cite{tv,tv2}. 
At present, the knowledge of such configurations is limited,
thus we will not conjecture about them.
In any case, for a cosmologically interesting warm inflation,
$N_e > 60$, without further assumptions,
the direct effect of defects at 
the onset will be negligible for post-inflation
magnetic field generation.

\subsection{Ferromagnetic Savvidy Vacuum Scenario}

In the Enqvist-Olesen ferromagnetic Savvidy vacuum scenario \cite{eo1},
the random variables are quark currents in the GUT-scaled 
radiation plasma at temperature $T$.
If a given quark traverses a distance $1/T$ without collision, it will
generate a magnetic field $\sim T^2$. This could help
to locally induce a transition
into the Savvidy vacuum at sufficiently high temperature 
$T^2 \gg B_{\rm min}$, where $B_{\rm min}$ is given in eq. (\ref{bmin}).
The Savvidy vacuum would act as a memory of the local currents. At very
high-T the quark currents generate very high magnetic fields, thus it
would be easy to reorient regions of Savvidy vacuum.   
For a plasma of quarks at a high temperature T, quarks will
criss-cross each other's paths significantly, which implies 
that a characteristic
coherence region for a local Savvidy vacuum bubble is
$ \sim 1/T^3$.  If these local bubbles were randomly distributed, then an
estimate of the rms-magnetic field at large scale would be possible with
magnitudes similar to those in the previous subsection.  This is the
scenario of magnetic fields that Enqvist and Olesen estimated.  Such a
scenario would work effectively during warm inflation.

However, a charged plasma tends to screen long range electromagnetic 
fields.  Thus charge and current distributions in a charged plasma 
are not random, especially local to any given charge.  Although 
rigorous calculations of magnetic screening in a charged non-Abelian
plasma have not been possible, evidence from analytic
\cite{lin} and numerical
calculations \cite{bls} suggest that a local magnetic field is screened 
at distances greater than $1/(g^2T)$.  Recently this point has been
examined in the context of the Enqvist-Olesen ferromagnetic Savvidy vacuum
scenario \cite{elpe}.  The conclusion in \cite{elpe} is that the charged
plasma is not an adequate source of large scale magnetic fields.
The largest scale magnetic field produced by the charged plasma 
$1/(g^2T)$ is very small compared to the characteristic scale of the
Savvidy vacuum,
which from eq. (\ref{bmin}) is
\begin{equation}
\Lambda_{SU(5)}=2 \times 10^{11}{\rm GeV}.
\label{lamsu5}
\end{equation}
Here, and throughout this
paper following \cite{elpe}, 
for SU(5) we set $\mu = M_{\rm GUT} = 10^{15} {\rm GeV}$
and $g(\mu) = 0.7$.  As such, the plasma 
fields do not have an adequate long range affect to 
induce a transition into the Savvidy
vacuum.

We observe that there is another source that can produce large scale
magnetic fields and at sufficient magnitude to induce the transition
into the ferromagnetic Savvidy vacuum, the monopoles that are produced at
the onset of symmetry breaking, which also is the onset of warm inflation.
Monopoles behave very differently from charged plasma particles as magnetic
field sources. They are very heavy and thus decorrelated from the motions in
the plasma. Although monopoles and antimonopoles are created
simultaneously, the Higg's gradients that initiate this process are
decorrelated. Thus at freeze-out there is no reason to expect the
monopole-antimonopole distribution to be so configured as to screen
magnetic fields at large scales. Thus the monopoles at freeze-out 
are a viable source of large scale magnetic fields that can induce
a transition of the gauge field system into the Savvidy vacuum.

Based on this observation, the following scenario suggests itself.
At $T_F \sim M_{GUT}$ a density of monopoles $n_M$ is created.
The local magnetic field $B_{\rm local} \approx n_M^{2/3},$
with corresponding coherence scale $n_M^{-1/3},$ will produce,
at the scale of the Hubble radius at the beginning of warm
inflation
\begin{equation}
H_{BI}^{-1}=\left(\sqrt{\frac{8\pi^3 g^*T_c^4}{45m_p^2}}
\right)^{-1},
\end{equation}
an rms-magnetic field from eq. (\ref{brmsgen}) of size
\begin{equation}
B_{rms}^{H_{BI}} \approx B_{\rm local} \sqrt{\frac{H_{BI}}{n_M^{1/3}}}
=\sqrt{n_M H_{BI}}.
\label{brms}
\end{equation}
To induce a transition to the Savvidy vacuum the source field must be 
the order of
or bigger than the magnetic field at the minimum of the ferromagnetic
effective potential
\begin{equation}
B_{rms} \stackrel{>}{\sim} B_{\rm min} \sim \frac{\Lambda^2}{g}
\sim 4 \times 10^{22} {\rm GeV}^2.
\label{brmsvalue}
\end{equation}

From eq. (\ref{brms}) this implies a monopole number density
\begin{equation}
n_M \geq \left( \frac{\Lambda^2}{g \sqrt{H_{BI}}} \right)^2
=2 \times 10^{34} {\rm GeV}^3.
\end{equation}
From the lower bound, as few as
$\sim 10$ monopoles
within the Hubble radius just before warm inflation
can induce the transition into the Savvidy vacuum.
However a rapid creation of the Savvidy vacuum may be necessary, before
full symmetry breaking has occurred.  In this case, a larger source
magnetic field is preferable, which in turn implies a monopole density
above this lower limit.

Although the monopoles solve the problem of a source field, a second
problem arises.
The non-Abelian effective potential eq. (\ref{effpotb}) 
was obtained from a
one-loop calculation
\cite{savvidy,masa,niol}
with a specific order of limits. If the large
distance limit is taken first and then the source magnetic field 
is removed, the effective potential has a new minimum at
the $B_{\rm min}$ in eq. (\ref{bmin}).  
An analogy to the thermodynamic limit has been given in
\cite{niol}, in which to obtain a nonvanishing magnetization, first the
infinite volume limit must be taken and then the external magnetic field
can be removed.  This underlies the need for an initial source magnetic
field at length scales larger than $> \Lambda^{-1}$. 
The validity of the one-loop approximation has been questioned in
\cite{maiani} on the basis that such classically unstable 
field configurations do not dominate the
functional integral and the problem is entirely nonperturbative.  On the
other hand, Nielsen and Olesen \cite{niol} have argued that the behavior
of the $\beta$-function eq. (\ref{betaf}) is sufficient for the
qualitative validity of the effective potential eq. (\ref{effpotb}).
They also have shown that the effective potential has an imaginary part
\begin{equation}
{\rm Im} {\rm V} = \frac{1}{8 \pi} \mu^4 
\exp{ \left( \frac{-96 \pi^2}{11Ng^2} \right)}
\label{decayp}
\end{equation} 
and identify this as the decay probability per unit spacetime volume.

They have interpreted the ferromagnetic vacuum as a metastable state,
below the perturbative vacuum, which decays to possible lower states and
eventually to the nonperturbative true non-abelian vacuum.
Based on this interpretation, the Savvidy vacuum induced by the monopoles
at the onset of warm inflation will decay. If it decays too soon before 
the inflationary period terminates, subsequent expansion will dilute the
magnetic fields produced up to then and they will not be cosmologically
interesting.  In warm inflation, observationally consistent expansion
e-folds requires an inflationary time period of order $100/H_{BI}$
or larger.  After this time interval, the decay probability per unit
three-volume from eq. (\ref{decayp}) is
\begin{equation}
\frac{P_{\rm decay}}{\rm volume} = \frac{100 \mu^4}{8 \pi H_{BI}}
\exp(\frac{-96 \pi^2}{11Ng^2})
= 2 \times 10^{34} {\rm GeV}^3.
\end{equation}
From this it follows that for Savvidy vacuum domains of radius 
$\sim (3 \times 10^{11} {\rm GeV})^{-1}$ or smaller after warm inflation,
the decay probability still is less than one.

For regions of Savvidy vacuum that have not decayed 
just after warm inflation, their effect will be imprinted on the plasma.
Since the plasma conserves flux \cite{parker,tuwi}, 
these magnetic fields will be
frozen into the plasma.  Right after warm inflation, 
the largest coherence scale for
magnetic fields made by the Savvidy vacuum will be
$\sim 10^{-12} {\rm GeV}^{-1}$, with magnitude 
from eq. (\ref{brmsvalue})
\begin{equation}
B^{EI} \approx B_{\rm min} = \frac{\Lambda^2}{g} \approx
10^{41} {\rm G}.
\label{bei} 
\end{equation}  
For comparison, in the warm inflation scenario considered above, the
temperature at the end of warm inflation is 
$T_{EI} \approx T_c/10 \approx M_{\rm GUT}/50 = 
2 \times 10^{13} {\rm GeV}$.
This implies the magnetic regions of Savvidy vacuum are much bigger that
the interparticle distance $1/T_{EI}$.  Note that the 
magnetic energy density of
the Savvidy vacuum $\rho_{B} \sim B_{\rm min}^2$ is much smaller than
either the radiation or vacuum energy during all of warm inflation, since
they are both $> T_{EI}^4$.  Also, there is no 
magnetic domain formation, since the Savvidy vacuum decays, after which,
the magnetic fields evolve with the rest of the plasma. 
If the magnetic fields eq. (\ref{bei}) 
subsequently follow the Hubble flow, the magnitude at time of
nucleosysthesis is $\sim 10^8$ G at the beginning
($T=1 {\rm MeV}$) and $\sim 10^4$G at the end ($T=0.01 {\rm MeV}$)
for coherent domains at these respective times of radius
smaller than $67 {\rm MeV}^{-1}$ and $6700 {\rm MeV}^{-1}$.
These magnitudes are below the big-bang nucleosynthesis bounds on magnetic
fields \cite{cheng}.
Similarly, the size of the magnetic field eq. (\ref{bei}) at the time
of galaxy formation (today) is $10^{-11}$ G at scales smaller than
$\sim 5$ cm. This implies from eq. (\ref{brms}) that at $\sim 100$kpc,
which is an interesting galactic scale, the rms-magnetic field is
$4 \times 10^{-23}$G \cite{vuconf}, 
which is a few orders of magnitude too small
to seed the galactic dynamo.
These estimates ignore nonlinear effects of
evolution after and even during warm inflation, such as inverse cascade 
\cite{beo,o1}, which would enhance the magnitude.

These estimates suggest favorable conditions for such a scenario. As such
it is worthwhile to examine arguments that could negate the scenario.
Firstly, scale considerations indicate that patches of Savvidy vacuum of
radius $ \sim \Lambda^{-1}_{\rm SU(5)}$ 
will coherently decay from the metastable
state, rather than say smaller patches in a longer time or visa-versa.
Since the domain size of Savvidy vacuum regions 
that are found to persist after warm
inflation (given above eq. (\ref{bei}))
are of order this characteristic length,
it is a borderline situation.  Secondly, although the Savvidy vacuum 
can be created in the symmetry restored regime,
the metastable state in the scenario must exist just below
the SU(5) symmetry breaking temperature, $T_c$.
Below $T_c$, twelve of the gauge fields have become massive, 
the X and Y bosons,
but they played an important role in establishing the Savvidy
vacuum.  Since the metastable state had already been created before these
bosons became massive, the question is how much does
the length and time scale for
decay of this state change from eq. (\ref{decayp})?
The SU(5) Savvidy vacuum may make its way to the SU(2) and SU(3) Savvidy
vacua below $T_c$, but due to the disparity in scales between these
phases, this fact does not help to quantify our question.  If one naively
changes N from 5 to 2 or 3 in eq. (\ref{decayp}), but still evaluates it
at scales of order $M_{\rm GUT}$, it would, in fact,
increase the lifetime of the
metastable state.

Finally, note that the basic features of this scenario are not specific to
warm inflation and also can be applied to supercooled inflation scenarios.
The essential requirement is that after the source magnetic field 
created by the monopoles is negligible, the inflationary epoch finishes
before the metastable Savvidy vacuum decays.  For a symmetry breaking warm
inflation scenario, this requirement is natural, since warm inflation
prefers small expansion e-folds, within a few orders of magnitude
above the lower observational bound.  Also, due to the presence of sizeable
radiation during warm inflation, the effects of the Savvidy vacuum
already are being imprinted on the plasma.  For a supercooled scenario,
a primary concern is that preheating/reheating may have adverse effects on 
an existing Savvidy vacuum.  The characteristic coherence scale during
reheating $\sim (M_{\rm GUT}/10)^{-1} - (M_{\rm GUT}/100)^{-1}$
is much smaller than $\Lambda^{-1}_{\rm SU(5)}$. Since reheating is a highly 
non-equilibrium process, this coherence scale will be the governing scale,
thus existing Savvidy vacua will likely be dismantled.

\section{Conclusion}

In this paper, we have examined GUT sources for magnetic fields and their
implications before and during a warm inflation regime.  
As a preliminary step, this has required a general determination of defect
density production during a symmetry breaking warm inflation scenario.
The thermal nature of the scenario implies this is determined at the onset
of the transition with a defect density no larger than the intrinsic scale
which at high temperature is set by $T \sim T_c$.  For SU(5) monopoles at
a number density $T_c^3$ at the onset of warm inflation, the Parker bound
and overdensity constraint require the number of e-folds
$N_e \geq 20$.  It has been noted that the initial density may be
substantially less based on the considerations by Zurek 
\cite{zurek,lagzur}.  

We have observed that the soon-to-be exiled 
SU(5) monopoles initially in the early
universe
may be good for something: sources of large scale magnetic
fields that can induce the gauge-field transition into the ferromagnetic
Savvidy vacuum.  A GUT warm inflation
scenario has been proposed for production of large
scale magnetic fields.
Although the monopoles solve the problem of an initial large scale
source, a second problem is the metastability of the Savvidy vacuum.  Due
to this, the inflationary epoch must be short, in order that the Savvidy
vacuum survive inflation and imprint its effect on the plasma.
We have found that the duration of warm inflation at the observational
lower bound $N_e \sim 60$ is of order the lifetime of the Savvidy
vacuum.  Some modifications to the scenario can improve this 
borderline situation.
For example, we estimated the source magnetic field from
the monopoles at the onset of warm inflation, but within
the first 5-10 e-folds of warm inflation it is sufficiently large to 
affect the Savvidy vacuum.  Likewise, if the Savvidy vacuum decays 
a few e-folds before warm inflation ends, there still
would be a sizeable imprint on the plasma leftover.  Irrespective of such
refinements, the strong exponential dependence in the Savvidy vacuum
results 
and its perturbative origin are the primary concerns.  

Accepting the
qualitative correctness of the one-loop Savvidy vacuum calculation, this
touchy quantitative issue can be avoided by increasing the disparity
between the lifetime of the Savvidy vacuum and the duration of warm
inflation.  This is possible at temperatures well above the GUT
scale $T \gg T_c$.  If this is a preinflation hot-big-bang regime,
the characteristic Savvidy vacuum region will be larger than the causal
horizon $1/H = t \approx m_p/(\sqrt{g*}T^2) \ll \Lambda^{-1}_{\rm
SU(5)}$.  In fact, very close to 
$m_p$, the causal horizon is smaller than the magnetic screening length
$1/H < 1/(g^2T)$.   

If the Savvidy vacuum is an acceptable solution in
this regime, the plasma particles could be the source to create it,
followed by a warm inflation regime.  The duration of an observationally
consistent warm inflation in this very high-T region is orders
of magnitude smaller than
the lifetime of the Savvidy vacuum, since the Hubble parameter is
much larger for $T \gg T_c$. As such, since
the Savvidy vacuum easily persists after
warm inflation, it will imprint its effect on the plasma. Furthermore,
now the magnetic screening effects work in favor of sustaining
the large scale fields, since the charged plasma is incapable of
reorienting the Savvidy vacuum, even if the temperature is very high.
To complete this scenario, the next step is the GUT transition.  To avoid
stable monopole production, flipped-SU(5) is 
appropriate if further inflation is
disallowed.
For the magnetic field scenario outlined above, 
cosmologically interesting magnetic fields are
created.
Since they are produced at an earlier time, assuming only Hubble flow
evolution, their magnitude today would be a few orders smaller than for
the scenario in subsection IVB.  As an aside, recall from the discussion
in subsection IVA that unstable
monopole-antimonopole configurations could be created at the flipped-SU(5)  
GUT transition. If so,  such configurations
could be an alternative and sole source for large scale magnetic
fields that induce the gauge field transition into the Savvidy vacuum.
This point will not be pursued further here.

Although the Savvidy vacuum scenario for magnetic field formation has
appealing features, it strongly relies on the quantitative properties of
the gauge field solution.  Unfortunately, to understand this beyond the perturbative
level probably will come in-hand with the understanding of QCD
confinement.  Nevertheless, the one-loop expressions are available, so it
is worth asking if they can be tested for the QCD Savvidy vacuum
\cite{vuconf}.
Neutron stars and heavy-ion collisions are cases in which Savvidy vacuum
bubbles may be created in the charged plasma, yet amongst a plethora of other
effects.  The conspicuous property about the Savvidy vacuum is
metastability, which may be exhibiting itself in resonance
states.   The lifetime of the $SU(3)$ Savvidy vacuum
$\sim \Lambda^{-1}_{QCD}$, is typical of several resonances.  The requirement
of a strong initial color-magnetic field source of magnitude 
$\Lambda^2_{\rm QCD} \approx (200 {\rm MeV})^2$
can easily be realized in local regions during a
high energy hadronic collision.  The cleanest process that contains
these two features is diffractive dissociation.  The pp-elastic scattering
slope parameter, which reflects the area of the scattered object is
$b  \approx 12 {\rm GeV}^{-2} \sim 1/\Lambda^2_{\rm QCD}$ \cite{dino}. 
When the Reggeon collides into such a
region on the hadronic disk, in some cases it could 
alter the local vacuum and place it in a metastable state, such as the
Savvidy vacuum.  Such a region would have small overlap with the rest of
the hadron, thus easily separate.   This picture to
the present elaboration is not very
specific to the Savvidy vacuum, beyond its metastablity
property.  Nevertheless, that is the only metastable
gauge-field vacuum that is known.

Excepting the reservations stated already, the Savvidy vacuum warm
inflation scenario involves generic features of gauge fields and
monopoles, which are commonplace in the present-day early universe picture.
No additional assumptions are made about field theory such as non-conformal
coupling \cite{tuwi,ratra,dolgov}. However if this effect was considered, it
would increase the size of the magnetic fields that are created, since
their redshift could be made to decrease slower than $1/R(t)^2$.

\section*{Acknowledgments}
We thank Tanmay Vachaspati for helpful discussions. AB also thanks
Per Elmfors,Kari Enqvist, 
Pablo Laguna, and Poul Olesen for helpful discussions.
This work was supported by the U. S. Department of Energy
under grant DE-FG05-85ER40226.
S.D.W. also gratefully acknowledges support from
the NASA/Tennessee Space Grant Consortium.

\end{document}